\documentclass[conference]{IEEEtran}
\IEEEoverridecommandlockouts
\usepackage{cite}
\usepackage{amsmath,amssymb,amsfonts}
\usepackage{algorithmic}
\usepackage{graphicx}
\usepackage{textcomp}
\usepackage{xcolor}
\def\BibTeX{{\rm B\kern-.05em{\sc i\kern-.025em b}\kern-.08em
    T\kern-.1667em\lower.7ex\hbox{E}\kern-.125emX}}
\begin{document}

\title{Smart Homes: Security Challenges and Privacy Concerns}

\author{\IEEEauthorblockN{Fraser Hall and Leandros Maglaras}
\IEEEauthorblockA{\textit{School of Computer Science and Informatics} \\
\textit{De Montfort University}\\
Leicester, UK}
\and
\IEEEauthorblockN{Theodoros Aivaliotis, Loukas Xagoraris and Ioanna Kantzavelou}
\IEEEauthorblockA{\textit{Dept. of Informatics and Computer Engineering} \\
\textit{University of West of Attika}\\
Athens, Greece}

}

\maketitle

\begin{abstract}
Development and growth of Internet of Things (IoT) technology has exponentially increased over the course of the last 10 years since its inception, and as a result has directly influenced the popularity and size of smart homes. In this article we present the main technologies and applications that constitute a smart home, we identify the main security and privacy challenges that smart home face and we provide good practices to mitigate those threats.  

\end{abstract}


\section{Introduction}

The smart home industry is growing in popularity in response to the exponential growth of the Internet of Things (IoT) market. Connected devices are becoming ever more popular and this trend is posing many security challenges. The cyber security industry has highlighted smart homes as one of the greatest concerns heading  into 2020, predicting the vast attack surface area they provide will be of great interest to malicious actors.

The purpose of this article is to perform a literature review surrounding these issues with focus surrounding the following statements:

\begin{itemize}
\item	Exponential growth in the IoT industry is responsible for the growth in popularity of smart homes

\item	Rapid growth of the smart home industry is raising serious concerns with regards to the security and privacy of its users

\item Security good practices must be used in all three phases of the lifecycle of a smart device; development, integration and usage of the IoT devices into a Smart Home
\end{itemize}

IoT devices and the idea of a smart home have been around for several years now; they are not new concepts. However, IoT technology continues to grow and develop and smart homes are becoming a reality for a greater number of people. This continuous expansion of technology and users leaves the fields of IoT and smart homes very much still in the Emerging Topic category. Looking into predicted cyber security trends for 2020, it is very apparent that many organisations believe the IoT and smart home industries to be ones of concern. Several different businesses have highlighted these as areas that will continue to grow and as a result will likely be targeted by attackers. Now more than ever there is a need to explore and address device weaknesses and user privacy concerns that surround the technology.

Towards IoT trends as expected in 2020, one was that Smart Home Devices will increase in Popularity. Also it was predicted that the growing charm of smart home devices using IoT would make it difficult to resist them even by those who first discarded the technology of smart homes \cite{zhang20202020}. The growing popularity and sheer numbers of different IoT devices that can now be integrated within the home has resulted in the sceptical accepting the idea that this technology will be a major part of the future. UbuntuPIT also investigated IoT trends that will shape the future and again concluded that Smart Home Demand Will Rise. Scholars believe that the smart home industry is driving the development of new technologies in the IoT space and as a result of this trend, soon  people will not direct the devices; instead, the devices will instruct the inhabitants of a smart home what they should do. A rather extreme and futuristic view, but it does show just how entwined the IoT industry and smart home industry are.

Expansion of IoT and smart home industries brings with it many security concerns. As part of their ‘5 Biggest Cybersecurity Trends In 2020 Everyone Should Know About’, Forbes identified ‘Data Theft Increases’ as one of their cyber security concerns.As stated in \cite{marr_2020} “Criminals have learned to piggyback into private networks through connected home appliances and smart devices, thanks to the lack of security standards among the thousands of device manufacturers and service providers”. Forbes believe that due to the presence of more, varied IoT devices within a greater number of people’s homes, there will be a much larger attack surface area for criminals to target, and as a result more people’s privacy could be at risk. TÜV Rheinland back up the thoughts of Forbes and several other organisations by reporting that “smart consumer devices are spreading faster than they can be secured”. They believe that “the number and performance of individual ‘smart’ devices is increasing every year, making them a very attractive target for cyber criminals. With the proliferation of smart devices, the attack surface could quickly increase hundreds or thousands of times” \cite{securitymagazinerss2020}.
 
\section {What is IoT?}

The Internet of Things (IoT) described in its simplest form, is the concept of connecting any device which has an on/off switch to the Internet and/or to other connected devices. These individual devices will have different purposes; collecting data about the way they are used or the environment around them and in turn sharing this information with other devices in the network. This technology is now applicable to an extraordinary number of varied devices, such as smart fridges, wearable tech and smart vehicles.

According to Cisco Internet Business Solutions Group (IBSG), the Internet of Things was “born” between 2008 and 2009, as this was the “point in time when more 'things or objects' were connected to the Internet than people". Quite staggeringly, the ratio of things to people grew from 0.08 in 2003 to 1.84 in 2010. Gartner reports that 2019 saw 14.2 billion connected devices, and they expect this to rise to 25 billion by 2021. 

\section{What is a Smart Home?}

According to the Department of Trade and Industry, a smart home is “a dwelling incorporating a communications network that connects the key electrical appliances and services, and allows them to be remotely controlled, monitored or accessed.” \cite{whatisasmarthome}. Here, the term remotely can refer to both within the dwelling and from outside the dwelling. Essentially, a smart home is a set of connected IoT devices that aim to enhance the living experience. These appliance and service devices can be categorised into six main areas:

\begin{enumerate}
\item Environmental – water meters, energy management, lighting
\item 	Security – alarms, cameras
\item	Home Entertainment – televisions, speakers
\item	Domestic Appliances – fridges, microwaves, coffee machines
\item	Information \& Communication – telephones, Internet
\item	Health – home assistance
\end{enumerate}

Every smart home will be set up according to an owner’s needs and wants, resulting in very different configurations. Devices can interact with each other in different ways, or sometimes not at all; it also means that devices can be connected to just one other device, or multiple devices. The flexibility that IoT devices provide to ensure users can implement home automation to suit them, while beneficial to the consumer, can lead to  security issues and privacy concerns.

\begin{figure*} 
\centering
\includegraphics[width=1.0\linewidth]{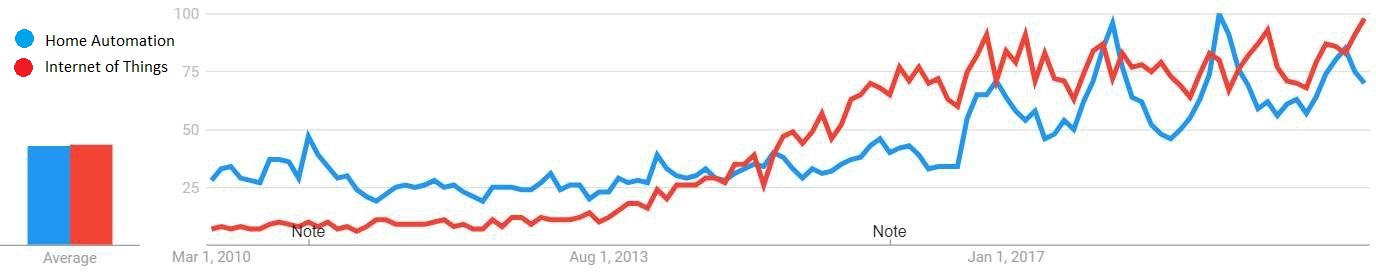}
\caption{Trends in ‘Home automation’ \& ‘Internet of things’ } \label{fig:1}
\end{figure*}

Google Trends \cite{googletrends} provides data on the popularity of searched terms and topics by users visiting their search engine. Data for ‘Home automation’ and ‘Internet of things’ Topics were compared to provide an insight into the reasoning behind this literature review. Figure \ref{fig:1} shows the interest level of users in the two topics over the last 10 years. Both topics have grown drastically in interest over the course of this time period as expected. However, another pattern has also been identified that proves relevant, and this is the change in trend that takes place between 2014 and 2017. IoT interest suddenly accelerates and jumps high above home automation, however a year later, home automation mirrors this sudden jump and falls back in line with IoT. This depicts the theory that the exponential growth of IoT devices is directly affecting the popularity of smart homes.

\begin{figure*} 
\centering
\includegraphics[width=1.0\linewidth]{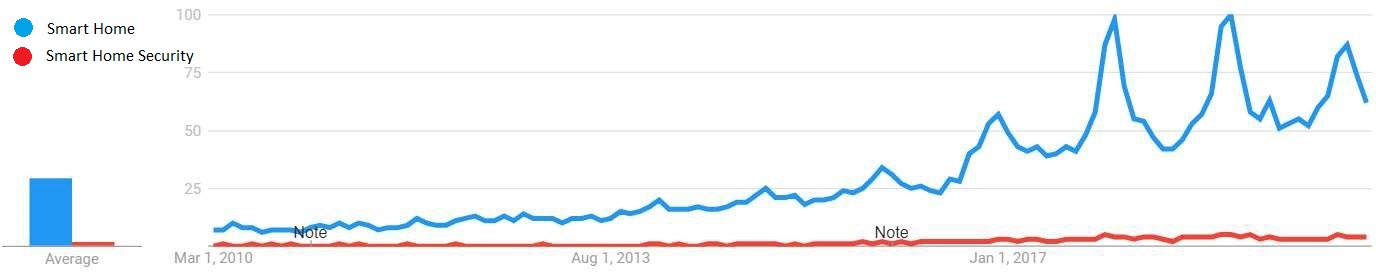}
\caption{Trends in ‘smart home’ \& ’smart home security’ } \label{fig:2}
\end{figure*}

Figure \ref{fig:2} compares the search terms ‘smart home’ and ‘smart home security’. The reasoning behind this was to identify whether users who are interested in implementing home automation technologies are also considering securing their devices. This is not a definitive statistic or clarification of a trend, simply an interest piece leading into the literature review. The results show that whilst the interest in smart homes has increased considerably over the last 10 years, the interest in securing them has not. This is cause for concern, as not all users will necessarily be technology minded, and as a result could be completely unaware of the security issues and privacy concerns that they are exposing themselves to.

\subsection{Connectivity inside a Smart Home}

The common point between Smart Home devices resides in the combination of “smartness” (data processing and connectivity) and the “local” nature of the use case (devices are in the user home). This means in practice that connectivity:
\begin{itemize}
    \item 	is always present in the devices, either limited to the Home Area Network or with access to the Internet.
\item	may be related to several kinds of communications protocols (direct, short-range or long-range, wired or wireless)
\item	may lead to several interconnected networks in the home and outside the home.
\end{itemize}

Such networks typically include:
\begin{itemize}
\item	One or several Home Area Networks (HAN), which are dedicated to local networks or subnetworks for Smart Home devices and sensors:
\item	One or several High Speed Networks, usually Wi-Fi networks, that may be provided by a set-top-box, mobile devices, etc.
\item	Personal Area Networks or ad-hoc networks created between several devices, for instance using low-speed connections (e.g. Bluetooth, ZigBee, etc.).
\item	Connections to Wide Area Networks (WAN):
\item	High Speed Networks, typically providing access to the Internet, for instance through the Internet Service Provider (ISP) network or the Mobile Network Operator (MNO) network.
\item	One or several Low Power Wide Area Networks (LPWAN), which provide WAN connectivity while requiring low power from the device (e.g. LoRaWAN [Long Range Wan], Sigfox, etc.).
\item	If the home uses a smart meter, this meter connects the home to the associated Advanced Metering Infrastructure (AMI) used to communicate with smart energy management devices.
\end{itemize}

It should be noted that real-life deployments of Smart Home might include only some of these networks, or might use them differently. Moreover, many elements of the Smart Home have connections to other domains: energy might have connections to the smart metering domain, devices related to assisted-living might have connections to the eHealth domain, etc. These connections might bring additional security constraints to these devices, notably in terms of compliance to national health or energy (critical infrastructure) requirements.

\subsection{Purpose of a Smart Home}

Development and growth of IoT technology has exponentially increased over the course of the last 10 years since its inception, and as a result has directly influenced the popularity and size of smart homes. Investigations into the changing smart home market identified that in 2015 it was worth an estimated $9.8$ million dollars and this was estimated to reach $43$ billion dollars in 2020 \cite{bugeja2016privacy}.  Research into this rapid growth in smart home popularity revealed 3 main driving factors, all of which are directly linked to IoT technology.

The purpose of a smart home is to create an environment where the inhabitants can live comfortably, with minimal effort \cite{ahvar2016analyzing}. Recent developments in IoT technology has provided additional functionality and practicality to the end user. When it comes to smart home setups, IoT technology is responsible for internet connectivity and remote management \cite{domb2019smart}. Without this functionality that provides the user with the ability to remotely control devices installed within their home, the popularity of smart devices implemented throughout the home would be far less than it currently is.

The second reason for the rapid growth in smart home popularity is safety. Smart homes are the most popular use for IoT devices, and they are used to provide safety and well-being for the users \cite{karimi2019smart}. Another study in 2016 showed that the smart home market was anticipated to double in the US with family safety being the greatest motivator. 

Finally, the popularity of smart homes has increased due to reduction in hardware costs \cite{anwar2017security}. When IoT devices first came out, the technology was reasonably expensive to acquire for the everyday user, certainly too costly to purchase several different devices to set up around the house. The development of IoT technology by a greater number of vendors and at a rapid rate has seen the availability of devices increase, and the price decrease. This has resulted in the technology becoming more affordable and attainable to a greater number of users.

\section{Security Challenges}

As with any rapidly growing industry, there are several challenges faced in the smart home area, most notably with the security of devices. Security is paramount to the success of the smart home. Feeling safe in your own home is a fundamental human need. The main security challenge posed by smart home setups is the nature of their network. Recent approaches to smart home implementations sees the interaction of several different devices, from multiple manufacturers in an ad-hoc manner.  This dynamic network setup opens the smart home to a greater threat landscape, with a much larger attack surface area. Devices will interact through different means, will transport and store data in different locations and will have different levels of security hardness \cite{sicari2018securing}.

Dynamic networks are not the only challenge posed by smart homes. IoT devices are extremely popular, and this growth in the industry has resulted in many start-up companies joining the market. Devices are now being developed and built quickly in order to profit from the demand, and this time pressure results in lower security standards being implemented during the build stage. Consequently, several devices are released which are hugely vulnerable and will provide an attacker with easy access to the rest of the more secure network that it  has been connected to.

Another challenge in the smart home area are the user themselves. Smart homes have become very popular over  a short period of time. Not all new users will be technology minded and security aware. Many users will simply connect devices to their network and begin to use them. As a result, there are many IoT devices connected that have default credentials or incorrect configurations, allowing easy unauthorised access.

\section{Privacy Issues}
The growth in popularity of smart homes has led to many security concerns in relation to scalability and interoperability issues amongst IoT devices.  Threats are rising exponentially in terms of frequency of attacks and complexity \cite{ferrag2018privacy}. However, it is not only security issues that interconnected home devices are exposing, there are also serious concerns regarding privacy. The nature of smart homes opens the user to many different privacy concerns from an unauthorised party obtaining unacceptable or inappropriate access to someone’s personal information, to psychological dimensions of privacy (solitude, reserve, isolation, anonymity, intimacy) or even a physical breach of privacy where the home itself is accessed by an unauthorised party \cite{heartfield2018taxonomy}. 

\subsection{Third Party Storage}

A big part in the development of the smart homes feature that allows remote access and monitoring was the introduction of cloud storage by third parties \cite{cook2018internet}. This allows data from your smart home to be accessible from anywhere. A worrying amount of personal data and private information could be stored by a third party. This was the case with Chinese company Orvibo who run an IoT management platform \cite{wang2020personal}. They were subject to a data breach that saw 2 billion records exposed in relation to smart home devices. Information such as passwords, account reset codes, precise geolocation and scheduling information was included in the breach. This would provide attackers with information on user routines and locations, potentially identifying when houses are empty, allowing burglary opportunities. The information could also render some of the devices useless, such as smart locks or security cameras, as attackers could now gain access to them.

\subsection{Secondary use of Data}

When purchasing and setting up new smart home devices, users may also be agreeing to allow their data to be used in a manner that isn’t solely for the device itself, as is the case with Amazon’s Alexa device. The device allows users to ask questions and interact with the device vocally. Voice samples submitted to the device are analysed by workers in order to improve Amazon’s voice recognition software. This poses a few concerns with data that may unintentionally be captured by the device. Private conversations had by users in their home may now not be so private. This use of data could also result in confidentiality and moral issues; two workers reported they had heard what they believe to have been a sexual assault when analysing voice samples \cite{day2019amazon}.

\subsection{Possible attacks}

The growth of smart homes and the diversity and volume of IoT devices connected within homes has increased the attack surface area for threat actors. As identified in Cyber Security Trends 2020, smart homes are going to be popular targets for malicious actors due to number of potential entry points. If an attacker can gain access to  a single smart device within the home, they could potentially have access to the entire network, resulting in the exposure of personal data and private information (eavesdropping attack). In 2017, an official watchdog in Germany instructed parents to destroy a talking doll called Cayla \cite{oltermann2017german}. It was discovered that the embedded Bluetooth device was insecure and could be exploited allowing an attacker to listen and talk to the child playing with it.

\section{Smart Televisions (TVs): A smart home privacy case study}

Any TV that can connect to the internet is a Smart TV and with technology where it is these days, it would be a struggle to buy a TV without this capability. When setting up a new TV, the user will be prompted to connect the device to the internet to ensure they receive the complete user experience; the ability to run apps, browse the internet, use streaming services. Manufacturers will also distribute patches and upgrades to their products and for these to be implemented automatically, it will need to have internet connectivity. Therefore, there are several benefits and legitimate reasons for TVs to be connected to the internet, however, opening this door has brought about a few privacy concerns \cite{rutledge2016privacy}.

In early 2017, Vizio landed in legal trouble as a result of findings by the Federal Trade Commission (FTC), and were fined $2.2$ million dollars. It was discovered that Vizio had been tracking the viewing habits of its customers through its smart TVs. On a second-by-second basis, Vizio collected a selection of pixels on the screen that it matched to a database of TV, movie, and commercial content.  This data was then sold on to third party companies for use for targeted advertising of customers. To make matters worse for Vizio, the data collecting feature was turned on by default and the customer was  not given  the option to opt out, so most  were unaware of the activity taking place.

It was also revealed in 2017 via WikiLeaks, that Samsung TVs had been subject to successful exploitation by the CIA (Confidentiality, Integrity, Availability). The exploit was dubbed ‘Weeping Angel’. This is an example of a physical attack against a smart home device, rather than the misuse of gathered data as explored previously. According to the information posted on WikiLeaks, a “Fake-Off" mode had been developed that turned off the screen and Light Emitting Diode (LED) status lights giving the illusion to the user that the TV was turned off \cite{hollister2017weeping}. However, the TV remote microphone was still switched on and voice conversations could be recorded. Weeping Angel was a huge breach of user privacy; any number of personal data points and private information could have been collected, dependent on what conversations took place in the vicinity of the TV remote.

 \section {Good practices for a secure Smart Home}
 
 In this section we provide a detailed list of security good practices to mitigate the threats identified in Smart Home Environment. We identify different types of good practices along with the dedicated countermeasures against given threats, for different types and classes of devices as well as for associated remote services.  They are separated into the three phases of the life cycle of devices and services:
 \begin {enumerate}
 \item	Development of Smart Home devices and services by device vendors and service providers. During this phase, the vendors and service providers define the requirements of the product, design, develop and test the product.
\item	Integration of devices by the end-user into his HAN. During this phase, the end-user configures and connects its Smart Home device to its HAN, potentially with support of the device vendor, the service provider, or the electronic communication provider.
\item	Usage of the devices and services until their end-of-life. Apart from direct and local interactions with his device, the end-user may also request support from the vendor and use on-line services related to the device through various communications channels. Thus, this phase may imply interactions with the device vendor, the service provider or the electronic communication provider for usage and decommission.
\end{enumerate}

\section{Discussion}

Development in technology and capabilities in the IoT industry have directly influenced the growth in popularity of smart homes. Increased functionality and internet connectivity providing remote accessibility and monitoring has been the main driving force behind this. Users have adopted the technology to increase comfort and ease of living, as well as improve security surrounding the home environment and loved ones.

This rapid increase in the adoption of technology and the inter-connectivity between different appliances has resulted in several security issues and privacy concerns. Users may not be aware to the full extent of the risks they are putting themselves at by adding devices to their home network that make a mundane daily task a little simpler. The nature of data collected by smart devices can reveal a lot about the user, whether it be personal information, daily routines or likes and dislikes. What a user once assumed to be confidential could now be used by third parties to push targeted advertising, track energy usage to see what time of day a user will be home or  in the case of Amazon Alexa, used to improve their own software and hardware. If a malicious actor were to gain access to this information, the consequences could be far worse and a lot more concerning as discovered with the Cayla doll in 2017.

The main security issues that Smart Home have to face are:
\begin{itemize}
\item	Not all Smart Homes are created equally. There are multiple design pathways that lead to functional smart homes, ranging between localized and integrated home-automation systems. These pathways have their own security and privacy peculiarities, but also have shared issues and vulnerabilities.
\item	Smart homes will have significantly privacy and data protection impacts. The increased number of interlinked sensors and activity logs present and active in the smart home will be a source of close, granular and intimate data on the activities and behavior of inhabitants and visitors. Also, the need for security in a Smart Home Environment is often underestimated. Current privacy regulations ensure that service providers will not intentionally collect private data. Smart Home actors comply with this regulation by privacy measures on the server-side of their services, which would arguably be enough in a world where no malicious actors were present. However, the absence of protection on the device-side means that private data collection might be relatively easy to perform on targeted individuals, even by attackers with low skills. Also, another major problem is that the asset owner is not aware of which private data could be leaked and how easy it is for an adversary to obtain these data.
\item	Several economic factors may lead to poor security in smart home devices. Companies involved in the smart home market include home appliance companies, small start-up companies, and even hobbyists. These groups are likely to lack security expertise, security budgets and access to security research networks and communities. Moreover, because the consumer market is cost-driven, with short time-to-market requirements, vendors lack incentives to enhance security in Smart Home devices and services.
\item	The interests of different asset owners in the smart home are not necessarily aligned and may even be in conflict. This creates a complex environment for security activity.
\item	Many IoT applications, Smart Home devices and services rely on other “building blocks”, which may cause unknown vulnerabilities to appear and be shared at large scale (in case these blocks have security flaws).
\item	Smart Home Environment result in new security challenges. Devices will have to meet higher privacy expectations than in usual IoT devices. These specifics lead to increased privacy risks for users, while the cost of keeping data safe might be too high for industry players.
\item	IoT devices in general are pervasive and dynamically interconnected. This has several consequences. Firstly, it increases the attack surface on a given device. Secondly, it increases the nuisance potential of a device after it has been compromised. Lastly, it increases the combinations between devices and services, leading to interoperability issues.
\end{itemize}

Smart home popularity and IoT devices will continue to grow in diversity and popularity over the coming years. The future could see a greater breach into the privacy of its users with the boundaries of legality already being pushed. Law enforcement agencies have started using smart home devices in court cases for data they can retrieve. However, manufacturers of these devices are still refusing to provide irretrievable requested data by law firms, ensuring their customers security and privacy remains intact. Kalev Leetaru has written an interesting piece on ‘What Happens When Our Smart Homes Enforce Moral Values?’. An interesting concept that worryingly may not seem too farfetched. He explores the notion that potentially a green energy supplier who is very conscious of climate change may start to dictate the usage of its energy by the customer rather than the other way around.

Smart homes will pay a valuable part in our future and the benefits and capabilities they provide will help push the boundaries of technology in the IoT industry. Security challenges need to be addressed with greater urgency and the notion of secure by design should be adopted by manufacturers. Privacy concerns need to be conveyed more clearly to users so they are aware of the exposure to the outside world their home could have. This awareness will hopefully result in users becoming more conscious of configurations and setups to ensure less vulnerabilities and weaknesses are added to what they believe to be the safety of their home network.

Future research on developing lightweight security mechanisms that will be suitable for devices with low power capabilities is needed. Moreover, novel secure authentication mechanisms that will be able to control access to complex environments like smart homes must be implemented. Encryption will continue to play a key role in securing smart homes and protecting their sensitive data. End to end encryption must be reassured especially when several smart devices with questionable level of security will be connected inside a smart home. Manufacturers need to put their users first and consider how secure their devices are.
Devices need to be designed with the concept of security in mind and cybersecurity certification schemes can play a key role in this. 
It is necessary to discuss and implement safeguards to protect users today from the potential threats of tomorrow.
 
\section{Conclusions}
The issue of cyber-security is much closer to Smart Home Environment than it has been usually understood. As a result of the increased usage of smart devices, people have been generating a huge amount of personal data and purse a great privacy risk to them, especially when this data is stored on small smart devices which are more vulnerable to privacy. This is mainly done without the knowledge of the user. Some of this information is been collected and processed by third parties and in some cases this is done without the user consent. What happens to collected data once it has been sourced from the end-user is also important. Whether the data has been hacked from an insecure system or device, or whether the user has been manipulated into giving up more data than he intended, the end result is the same. The sourced data can be repurposed by the companies who collected it, or sold on to third parties, and the user does not have control over this, or in some cases even knowledge of it. 

Any precaution implemented at the user end can only limit the amount of data collected. For proper security, and privacy, these issues need to be addressed by the manufactures of smart devices and the designers of their interfaces. In a Smart Home, instead of having to rely on the user to check the security level of the smart devices, the onus should be on the manufacturer. Value sensitive design is a way to protect users from unwanted consequences of using technology. General Data Protection 
Ethical design with the user’s best interest in mind is needed – avoiding dark patterns and offering users simple straightforward options regarding their security and privacy.

\bibliographystyle{IEEEtran}
\bibliography{biblio}

\begin{thebibliography}{10}
\providecommand{\url}[1]{#1}
\csname url@samestyle\endcsname
\providecommand{\newblock}{\relax}
\providecommand{\bibinfo}[2]{#2}
\providecommand{\BIBentrySTDinterwordspacing}{\spaceskip=0pt\relax}
\providecommand{\BIBentryALTinterwordstretchfactor}{4}
\providecommand{\BIBentryALTinterwordspacing}{\spaceskip=\fontdimen2\font plus
\BIBentryALTinterwordstretchfactor\fontdimen3\font minus
  \fontdimen4\font\relax}
\providecommand{\BIBforeignlanguage}[2]{{%
\expandafter\ifx\csname l@#1\endcsname\relax
\typeout{** WARNING: IEEEtran.bst: No hyphenation pattern has been}%
\typeout{** loaded for the language `#1'. Using the pattern for}%
\typeout{** the default language instead.}%
\else
\language=\csname l@#1\endcsname
\fi
#2}}
\providecommand{\BIBdecl}{\relax}
\BIBdecl

\bibitem{zhang20202020}
X.~Zhang and W.~T. Yue, ``A 2020 perspective on “transformative value of the
  internet of things and pricing decisions”,'' \emph{Electronic Commerce
  Research and Applications}, p. 100967, 2020.

\bibitem{marr_2020}
\BIBentryALTinterwordspacing
B.~Marr, ``The 5 biggest cybersecurity trends in 2020 everyone should know
  about,'' Jan 2020. [Online]. Available:
  \url{https://www.forbes.com/sites/bernardmarr/2020/01/10/the-5-biggest-cybersecurity-trends-in-2020-everyone-should-know-about/}
\BIBentrySTDinterwordspacing

\bibitem{securitymagazinerss2020}
\BIBentryALTinterwordspacing
``Top seven cybersecurity trends in 2020,'' Feb 2020. [Online]. Available:
  \url{https://www.securitymagazine.com/articles/91696-top-7-cybersecurity-trends-in-2020}
\BIBentrySTDinterwordspacing

\bibitem{whatisasmarthome}
\BIBentryALTinterwordspacing
``What is a "smart home"?: Smart home energy.'' [Online]. Available:
  \url{http://smarthomeenergy.co.uk/what-smart-home}
\BIBentrySTDinterwordspacing

\bibitem{googletrends}
\BIBentryALTinterwordspacing
 [Online]. Available: \url{https://trends.google.com/trends/}
\BIBentrySTDinterwordspacing

\bibitem{bugeja2016privacy}
J.~Bugeja, A.~Jacobsson, and P.~Davidsson, ``On privacy and security challenges
  in smart connected homes,'' in \emph{2016 European Intelligence and Security
  Informatics Conference (EISIC)}.\hskip 1em plus 0.5em minus 0.4em\relax IEEE,
  2016, pp. 172--175.

\bibitem{ahvar2016analyzing}
E.~Ahvar, N.~Daneshgar-Moghaddam, A.~M. Ortiz, G.~M. Lee, and N.~Crespi, ``On
  analyzing user location discovery methods in smart homes: A taxonomy and
  survey,'' \emph{Journal of Network and Computer Applications}, vol.~76, pp.
  75--86, 2016.

\bibitem{domb2019smart}
M.~Domb, ``Smart home systems based on internet of things,'' in \emph{Internet
  of Things (IoT) for Automated and Smart Applications}.\hskip 1em plus 0.5em
  minus 0.4em\relax IntechOpen, 2019.

\bibitem{karimi2019smart}
K.~Karimi and S.~Krit, ``Smart home-smartphone systems: Threats, security
  requirements and open research challenges,'' in \emph{2019 International
  Conference of Computer Science and Renewable Energies (ICCSRE)}.\hskip 1em
  plus 0.5em minus 0.4em\relax IEEE, 2019, pp. 1--5.

\bibitem{anwar2017security}
M.~N. Anwar, M.~Nazir, and K.~Mustafa, ``Security threats taxonomy: smart-home
  perspective,'' in \emph{2017 3rd International Conference on Advances in
  Computing, Communication \& Automation (ICACCA)(Fall)}.\hskip 1em plus 0.5em
  minus 0.4em\relax IEEE, 2017, pp. 1--4.

\bibitem{sicari2018securing}
S.~Sicari, A.~Rizzardi, D.~Miorandi, and A.~Coen-Porisini, ``Securing the smart
  home: A real case study,'' \emph{Internet Technology Letters}, vol.~1, no.~3,
  p. e22, 2018.

\bibitem{ferrag2018privacy}
M.~A. Ferrag, A.~Derhab, L.~Maglaras, M.~Mukherjee, and H.~Janicke,
  ``Privacy-preserving schemes for fog-based iot applications: Threat models,
  solutions, and challenges,'' in \emph{2018 International Conference on Smart
  Communications in Network Technologies (SaCoNeT)}.\hskip 1em plus 0.5em minus
  0.4em\relax IEEE, 2018, pp. 37--42.

\bibitem{heartfield2018taxonomy}
R.~Heartfield, G.~Loukas, S.~Budimir, A.~Bezemskij, J.~R. Fontaine,
  A.~Filippoupolitis, and E.~Roesch, ``A taxonomy of cyber-physical threats and
  impact in the smart home,'' \emph{Computers \& Security}, vol.~78, pp.
  398--428, 2018.

\bibitem{cook2018internet}
A.~Cook, M.~Robinson, M.~A. Ferrag, L.~A. Maglaras, Y.~He, K.~Jones, and
  H.~Janicke, ``Internet of cloud: Security and privacy issues,'' in
  \emph{Cloud Computing for Optimization: Foundations, Applications, and
  Challenges}.\hskip 1em plus 0.5em minus 0.4em\relax Springer, 2018, pp.
  271--301.

\bibitem{wang2020personal}
Z.~Wang, ``Personal information security risks and legal prevention from the
  perspective of network security,'' in \emph{The International Conference on
  Cyber Security Intelligence and Analytics}.\hskip 1em plus 0.5em minus
  0.4em\relax Springer, 2020, pp. 113--117.

\bibitem{day2019amazon}
M.~Day, G.~Turner, and N.~Drozdiak, ``Amazon workers are listening to what you
  tell alexa,'' \emph{Bloomberg, April}, 2019.

\bibitem{oltermann2017german}
P.~Oltermann, ``German parents told to destroy doll that can spy on children,''
  \emph{The Guardian}, 2017.

\bibitem{rutledge2016privacy}
R.~L. Rutledge, A.~K. Massey, and A.~I. Ant{\'o}n, ``Privacy impacts of iot
  devices: a smarttv case study,'' in \emph{2016 IEEE 24th International
  Requirements Engineering Conference Workshops (REW)}.\hskip 1em plus 0.5em
  minus 0.4em\relax IEEE, 2016, pp. 261--270.

\bibitem{hollister2017weeping}
S.~Hollister, ``Weeping angel: Did the cia really hack into tvs,''
  \emph{Retrieved from}, 2017.

\end{thebibliography}
\end{document}